\documentclass[
]{ceurart}

\usepackage{listings}
\usepackage{url}
\begin{document}

\copyrightyear{2025}
\copyrightclause{Copyright for this paper by its authors.
  Use permitted under Creative Commons License Attribution 4.0
  International (CC BY 4.0).}

\conference{RCIS 2025 Research Projects Track}

\title{Low-code to fight climate change: the Climaborough project}


\author[1,2]{Aaron Conrardy}[%
orcid=0000-0002-3030-4529,
email=aaron.conrardy@list.lu,
url=https://aaronconrardy.com,
]
\cormark[1]
\fnmark[1]
\address[1]{Luxembourg Institute of Science and Technology, Belval, Luxembourg}
\address[2]{University of Luxembourg, Belval, Luxembourg}
\address[3]{ANCI Toscana, Florence, Italy}
\address[4]{Daten-Kompetenzzentrum für Städte und Regionen (DKSR) GmbH, Berlin, Germany}

\author[1]{Armen Sulejmani}[%
orcid=0009-0005-9571-8780,
email=armen.sulejmani@list.lu,
url=https://armen-sulejmani.com,
]
\fnmark[1]
\cormark[1]

\author[1]{Cindy Guerlain}[%
orcid=0000-0002-2617-4434,
email=cindy.guerlain@list.lu
]

\author[1]{Daniele Pagani}[%
orcid=0009-0004-1273-8691,
email=daniele.pagani@list.lu
]

\author[3]{David Hick}[%
email=david.hick@dksr.city,
orcid=0009-0007-6304-3189
]

\author[4]{Matteo Satta}[%
email=m.satta@matteosatta.com
]

\author[1,2]{Jordi Cabot}[%
orcid=0000-0003-2418-2489,
email=jordi.cabot@uni.lu,
url=https://jordicabot.com,
]

\cortext[1]{Corresponding author.}
\fntext[1]{These authors contributed equally.}

\begin{abstract}
  The EU-funded Climaborough project supports European cities to achieve carbon neutrality by 2030. Eleven cities in nine countries will deploy in real conditions products and services fostering climate transition in their local environment. The Climaborough City Platform is being developed to monitor the cities' overall progress towards their climate goals by aggregating historic and real-time data and displaying the results in user-friendly dashboards that will be used by non-technical experts to evaluate the effectiveness of local experimental initiatives, identify those that yield significant impact, and assess the potential consequences of scaling them up to a broader level.  

  In this paper, we explain how we have put in place a low-code/no-code strategy in Climaborough in response to the project's aim to quickly deploy climate dashboards. A low-code strategy is used to accelerate the development of the dashboards. The dashboards embed a no-code philosophy that enables all types of citizen profiles to configure and adapt the dashboard to their specific needs.  
\end{abstract}


\maketitle


\section{Introduction}
Climaborough\footnote{\url{https://climaborough.eu/}}\footnote{\url{https://cordis.europa.eu/project/id/101096464}} is a research project, co-funded by the European Union and CINEA, aimed at bridging the gap between the design and implementation of urban innovations, tackling climate change and its consequential need for rapid adaptation and mitigation. Specifically, it aims to overcome the bottlenecks when transitioning from prototyping to testing and to market deployment of innovation. It is a four-year project started on January 1, 2023 and coordinated by ANCI Toscana\footnote{\url{https://ancitoscana.it/}} with the participation of 27 additional partners, of which 14 European cities engaged in their ecological and digital transition.
Seven work packages (WP) were defined to tackle different tasks: (WP1) Urban Planning and Climate Neutrality Evaluation, (WP2) Climaborough City Platform, (WP3) CLIMHUBS Setup, Co-Creation and Collaboration, (WP4) Innovative Procurement, (WP5) Climate Sandbox Demonstration in Real Environments, (WP6) Dissemination, Communication and Exploitation and (WP7) Management.

The primary outcome of the project will be the development of a structured process incorporating a set of tactical tools designed in collaboration with domain experts, including an innovative procurement process aimed at accelerating cities' capacity to implement climate transition strategies within urban planning. 
More specifically, as part of the process, cities are engaged in defining their specific needs, which are addressed through an innovative procurement process managed by ANCI Toscana. This process enables cities to identify and adopt disruptive solutions across various sectors, including energy, mobility, waste management, and circular economy. These solutions are subsequently implemented and tested using a sandbox methodology, and the resulting data is integrated into the platform. The impact of these initiatives is then assessed through a Climate Neutrality Framework\footnote{\url{https://ec.europa.eu/research/participants/documents/downloadPublic?documentIds=080166e517cb7a44&appId=PPGMS}}, which facilitates the estimation of their broader scalability and effectiveness and helps cities in their decision to adopt them or not at scale.

To evaluate the progress and effectiveness of the solutions, and the project in general, it is critical to put in place an infrastructure to make sure the generated data (at the specific solution level, at the city level, at the project level...) will be monitored and evaluated. This includes defining (1) KPIs related to the solution specific goals, (2) Metrics to estimate the solution's contribution to overarching climate KPIs, and (3) KPIs describing the overall cities' progress towards climate related goals (such as reaching zero carbon emission) using aggregated data from solutions and beyond.

To help cities assess the effectiveness of implemented solutions and better understand their potential impact on achieving climate neutrality on a larger scale, the Climaborough City Platform will be developed by WP2, led by the Daten-Kompetenzzentrum für Städte und Regionen\footnote{\url{https://www.dksr.city}} (DKSR) in collaboration with the Luxembourg Institute of Science and Technology\footnote{\url{https://www.list.lu/}} (LIST) and the Institut Mines-Télécom\footnote{\url{https://www.imt.fr/}} (IMT), and will incorporate a data ingestion pipeline, dashboards for visualizing results and a digital twin component.

The creation of the dashboards is especially important and challenging, as it needs to cater to a variety of users and enable the adaptation of the core dashboard to the specific data and solutions under evaluation in a given city. To make things worse, this adaptation sometimes needs to be done by non-technical people in the public administration, as they rely on the visualization to support their decision-making. 

In this sense, this paper focuses on describing how the project has followed a low-code and no-code strategy \cite{coin,cabot2024thelowcode} to create these flexible dashboards in an optimal way, highlighting the benefits of this type of technologies in complex data manipulation and visualization scenarios such as that of this project. In short, the low-code part is used to speed up the development of the core dashboard components while a no-code mechanism, embedded in the generated dashboards, allows users to add, remove and configure the dashboard widgets. This strategy is implemented on top of the low-code platform BESSER\footnote{\url{https://github.com/BESSER-PEARL/BESSER}} \cite{alfonso2024building}. Therefore, we typically refer to these dashboards as BESSER-dashboards.


Next sections are structured as follows. Section \ref{section:overview} describes the Climaborough City Platform and its application in Climaborough.
Then, Section \ref{section:besser} provides more details on the low-code respectively no-code contribution. 
Afterwards, Section \ref{section:current} presents the current state of the platform and discusses some of the choices we made based on encountered challenges.
Finally, Section \ref{section:conclusion} concludes this paper and presents the next steps.

\section{The Climaborough City Platform}\label{section:overview}


The BESSER-dashboards are part of the Climaborough City Platform. This platform is a data-driven system for monitoring and evaluating the effectiveness of urban climate solutions, while tracking progress toward the cities' climate transition goals. It serves as a centralized data aggregator, integrating streams from heterogeneous sources. By consolidating these diverse inputs, the platform provides a comprehensive view of key performance indicators through its dashboards.

Platform requirements were defined through co-creation workshops and interviews with cities and experts, in which predefined features were ranked based on their priority by the participants to ensure the functional and non-functional needs were met.

\begin{figure}
    \centering
    \includegraphics[width=0.9\linewidth]{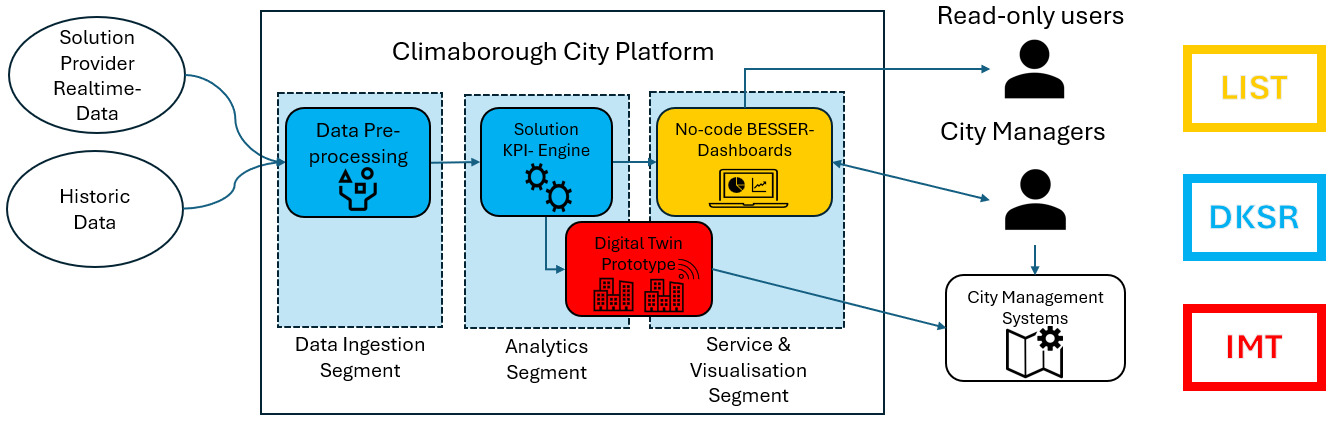}
    \caption{Climaborough City Platform overview}
    \label{fig:climaplatoverview}
\end{figure}

In the following, we will outline the architecture of the Climaborough City Platform, as illustrated in Figure \ref{fig:climaplatoverview}.

The Data Ingestion Segment is responsible for gathering and managing data from diverse sources. It integrates historical records, real-time data provided by solution providers, and additional datasets from external climate initiatives such as Copernicus\footnote{\url{https://www.copernicus.eu/en}}. This layer relies on traditional data collection and processing techniques, ensuring robustness and reliability. 

Once data is ingested, it is processed in the Analytics Segment, where it is aggregated and analyzed based on Key Performance Indicators (KPIs). These KPIs, developed in collaboration with cities and domain experts, allow stakeholders to measure the effectiveness of their climate strategies.

Additionally, this segment integrates a Digital Twin (DT), proposed by the IMT, which enables predictive modeling and scenario analysis. The DT leverages real-time and historical data to simulate the potential impacts of various climate strategies, helping cities anticipate challenges and optimize their policies accordingly.

The final layer translates the processed data into actionable insights through visualizations on a dashboard that follows a no-code approach. Its primary goal is to allow cities to interactively explore and visualize climate-related data and KPIs without requiring technical expertise. The no-code interface enables users to create and modify dashboards with a simple drag-and-drop mechanism, significantly lowering the barrier to data-driven decision-making.

This part is where the platform’s low-code and no-code innovations take effect. In the following sections, we will explore how this approach enables rapid dashboard deployment, enhances usability and adoption, and empowers cities in their climate transition efforts.

        


\section{BESSER-dashboards: Low-code approach to configure no-code dashboards}\label{section:besser}

\begin{figure}
    \centering
    \includegraphics[width=\linewidth]{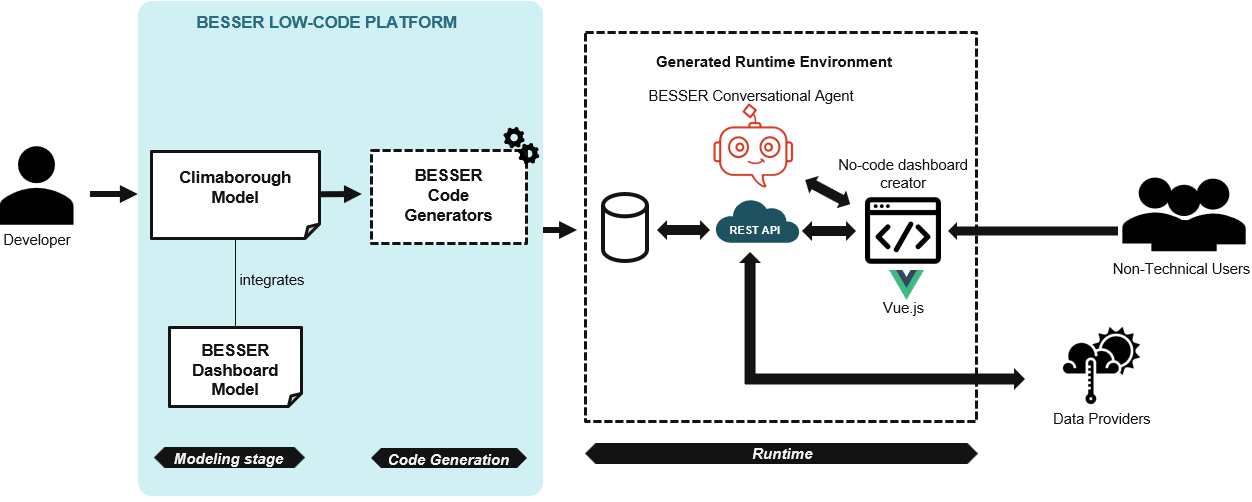}
    \caption{Overview of BESSER-dashboards integration in Climaborough}
    \label{fig:enter-label}
\end{figure}

Our main objective with the BESSER-dashboards is to speed up the development of dashboards and their integration in data-driven projects while guaranteeing that the resulting dashboards can be created and adapted by non-technical stakeholders. 
As illustrated in Figure \ref{fig:enter-label}, our approach is built around BESSER, a robust low-code platform that guides developers through two primary stages: modeling and code generation.

The creation of a final dashboard involves then two stages. In the first one, technical people select the data and the core features of the dashboard. In a second stage, non-technical people adapt the dashboard to their specific needs. Let us see both phases in more detail.

During the modeling stage, designers and developers define a data model that forms the foundation of their application. While data modeling still needs to be led by technical people, the abstract nature (thanks to the use of graphical modeling languages) still allows for collaboration with non-technical stakeholders, allowing for their involvement at an early stage. This model is then fed directly into our automated code generators, which produce a complete backend and frontend environment for a web application consisting of the no-code dashboards and additional data management features. This process eliminates repetitive manual coding and ensures that the generated code is consistent, scalable, and tightly aligned with the specified data structure. In situations where a project already has an existing backend, BESSER can generate only the necessary additional components, ensuring seamless integration without requiring a complete system overhaul. This web application comes with a dashboard front-end preconfigured, via the use of low-code techniques, to be aware of the data model and include all the necessary connections to read the data from the backend.

After this initial configuration, non-technical users are now allowed to modify all dashboard widgets. BESSER-dashboards embed a no-code philosophy, enabling diverse users such as city planners or sustainability officers to drag-and-drop widgets for tailored views. Users can select a datasource from a list and associate it with a visualization through simple drag-and-drop actions, eliminating the need for manual coding or technical expertise. The visualization is automatically configured based on the datasource schema for optimal clarity. Additionally, a conversational agent further simplifies interactions, allowing users of all technical background to adjust their dashboard or query data using natural language.

\begin{figure}
    \centering
    \includegraphics[width=0.7\linewidth]{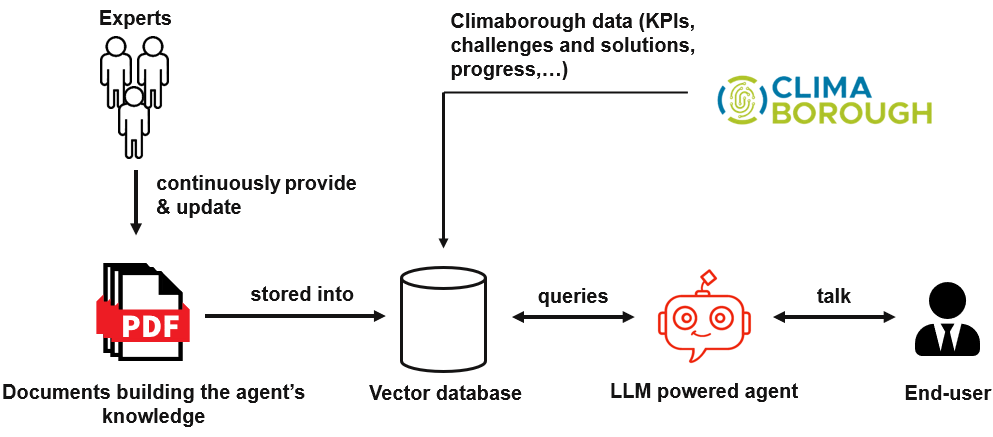}
    \caption{Overview of pipeline to create RAG powered Agent}
    \label{fig:rag}
\end{figure}

In addition to the basic dashboard interaction, an AI-powered multilingual conversational agent is also available on the dashboard. This agent plays two different roles:
\begin{enumerate}
    \item Help in the no-code strategy by offering a conversational (textual and audio) dashboard interaction to let users create, modify and read dashboard content by directly chatting with the agent.  
    \item Answer data questions. Complementing the visualizations, cities can also ask the agent questions. Thanks to its curated internal knowledge and the use  of Retrieval-Augmented Generation (RAG), which integrates Large Language Models (LLMs), the agent is able to provide useful answers. The knowledge body is being created in collaboration with domain experts from other work packages, while also including research and results from other climate initiatives such as NetZeroCities\footnote{\url{https://netzerocities.eu}}, and Climaborough project deliverables and data. This facilitates both the knowledge transfer from experts and replication of implemented solutions, given that many of the challenges cities face are not unique. This process is depicted in Figure \ref{fig:rag}.

\end{enumerate}



\section{Current state of the Climaborough City Platform}\label{section:current}

\begin{figure}
    \centering
    \includegraphics[width=\linewidth]{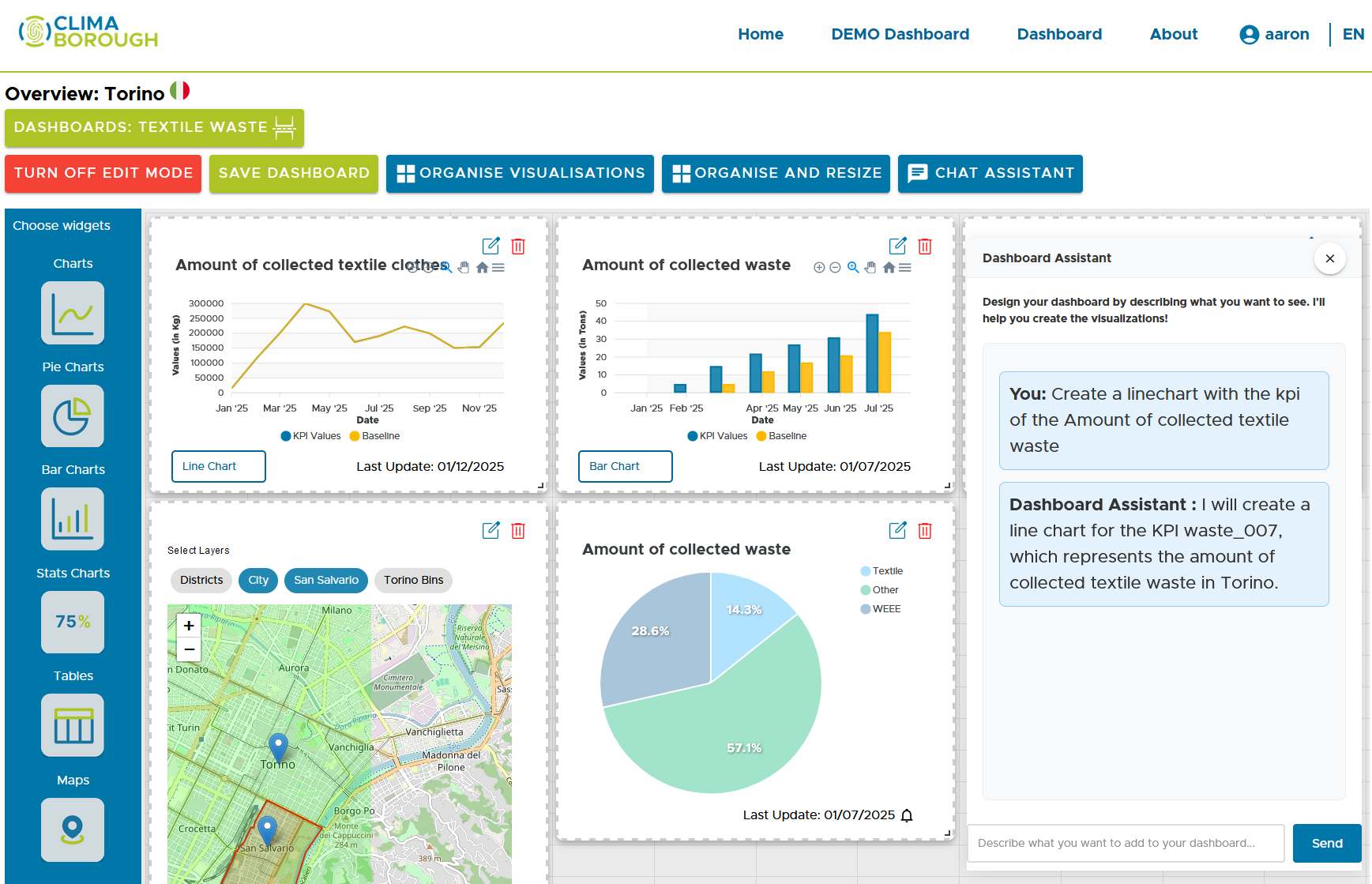}
    \caption{Screenshot of the dashboard editor, showcasing the drag-and-drop interface alongside the agent for creating dashboards}
    \label{fig:builder}
\end{figure}


An early version of a data processing module is under development, exploring different technologies to streamline data ingestion and transformation. This module is currently tested with various approaches, including direct integration from source like Google Drive to API Endpoints. To improve data accessibility and organization, the development of a data catalog is being considered

At the same time, the BESSER-dashboards leverage a structured backend and an interactive frontend that streamline the dashboard creation. The backend leverages a model-driven approach, enabling the automatic generation of a REST API and a database schema directly from the defined metamodel, ensuring consistency between data structures and API endpoints. On the frontend, the no-code interface provides already significant flexibility and customization possibilities, as users can drag and drop visual components, position and resize them dynamically and adapt the design and descriptions of the visualizations (e.g., changing the color of a visualization or the title). The low-code process is already in place, and in collaboration with the cities, we have started modeling the data models corresponding to their needs. Figure \ref{fig:builder} contains a screenshot of the dashboard creator, where one can see the possible visualizations that the dashboard supports.

The conversational agent-based dashboard creation feature further streamlines the process by allowing users to generate and modify dashboards through natural language commands. Currently in an advanced prototyping phase, efforts are focused on enhancing contextual understanding and enabling more complex customization options.
Additionally, the RAG features are in place, which provide structured climate-related insights. Ongoing improvements aim to refine response accuracy and expand document coverage with expert contributions.




\subsection{Discussion}\label{section:discussion}

In this section, we cover some reflections from the use of information system techniques and in particular low-code / no-code techniques in the context of the project.

\paragraph{Flexibility of model-based and low-code approaches}

One of the initial choices we considered was using an existing dashboard tool for the Climaborough City Platform (e.g. Grafana). However, after careful evaluation, we decided to develop our own solution to maintain full control over its features, customization, and adaptability. Moreover, thanks to following a model-based approach, we could easily change the tech stack targeted by the code generators in case cities had some preferences / restrictions in the type of infrastructure they wanted to use. Given the evolving requirements of a research project involving so many partners, this was a critical concern that low-code helps us to address.

\paragraph{Balancing low-code and no-code approaches}

Another crucial consideration was whether to implement a fully low-code or no-code solution instead of a mixed one as we have done in the project. A fully no-code approach would have introduced many limitations as users would have been restricted to a predefined number of templates to build the complete dashboard. A fully low-code one would have offer full flexibility but required technical expertise during the customization phase. Our mixed approach aims to combine both worlds. 


\paragraph{Need for AI-enhanced components}
While our dashboard is built on core information systems and model-based techniques, it was inevitable to add also AI techniques in the mix. Users expect them (e.g. in the form of a conversational agent). Therefore, it is clear that, more and more, we need to combine classical engineering techniques with AI ones to build smart software systems \cite{CabotC23}. Fortunately, BESSER is already created with this goal in mind and it was therefore easy to integrate conversational agent development in our low-code process.


\paragraph{Early validation}

Preliminary testing by our project partners has demonstrated significant interest in the dashboard’s core features. Cities have responded positively to the intuitive nature of the no-code interface, highlighting its ease of use and accessibility. The positive feedback from partner cities highlights how blending low-code for setup with no-code for user interaction makes climate data visualization more accessible and engaging. Cities have embraced the ability to effortlessly shape their own dashboards, giving them direct control over their data without needing technical expertise. This hands-on approach fosters greater involvement and ownership in their climate monitoring efforts.









\section{Conclusion and future work}\label{section:conclusion}

In an effort to support the performance tracking of the pilots deployed in the Climaborough project, we presented the concept of the Climaborough City Platform.
Specifically, given the project's goal of fast implementation of climate solutions, we focused on the low-code approach to quickly configure the backend and frontend environment of a no-code dashboard creator for non-technical users, enhanced by advanced interaction methods.

Based on the feedback provided by cities, we will improve both aspects. We also plan to study in what other parts of the project architecture (potentially involving other WPs), a low-code strategy could also make sense. For instance, in the creation and execution of climate simulation scenarios. 


Finally, we plan to connect the Climaborough City Platform to other climate initiatives such as the NetZeroCities platform\footnote{\url{https://netzerocities.eu/}} or the Climaborough twin project UP2030\footnote{\url{https://cordis.europa.eu/project/id/101096405}}. This could include pushing or pulling data from the NetZeroCities knowledge repository or including NetZeroCities benchmarks.

\begin{acknowledgments}
This project is supported by the Luxembourg National Research Fund (FNR) PEARL programme under the grant agreement 16544475 and the Climaborough project, co-funded by the European Union under the grant agreement 101096464. 
\end{acknowledgments}

\bibliography{sample-ceur}

\begin{thebibliography}{4}
\expandafter\ifx\csname natexlab\endcsname\relax\def\natexlab#1{#1}\fi
\providecommand{\url}[1]{\texttt{#1}}
\providecommand{\href}[2]{#2}
\providecommand{\path}[1]{#1}
\providecommand{\DOIprefix}{doi:}
\providecommand{\ArXivprefix}{arXiv:}
\providecommand{\URLprefix}{URL: }
\providecommand{\Pubmedprefix}{pmid:}
\providecommand{\doi}[1]{\href{http://dx.doi.org/#1}{\path{#1}}}
\providecommand{\Pubmed}[1]{\href{pmid:#1}{\path{#1}}}
\providecommand{\bibinfo}[2]{#2}
\ifx\xfnm\relax \def\xfnm[#1]{\unskip,\space#1}\fi
\bibitem[{Di~Ruscio et~al.(2022)Di~Ruscio, Kolovos, Lara, Pierantonio, Tisi, and Wimmer}]{coin}
\bibinfo{author}{D.~Di~Ruscio}, \bibinfo{author}{D.~Kolovos}, \bibinfo{author}{J.~Lara}, \bibinfo{author}{A.~Pierantonio}, \bibinfo{author}{M.~Tisi}, \bibinfo{author}{M.~Wimmer},
\newblock \bibinfo{title}{Low-code development and model-driven engineering: Two sides of the same coin?},
\newblock \bibinfo{journal}{Software and Systems Modeling} \bibinfo{volume}{21} (\bibinfo{year}{2022}).
\bibitem[{Cabot(2024)}]{cabot2024thelowcode}
\bibinfo{author}{J.~Cabot}, \bibinfo{title}{The low-code handbook: Learn how to unlock faster and better software development with low-code solutions}, \bibinfo{publisher}{Jordi Cabot}, \bibinfo{year}{2024}. \URLprefix \url{https://lowcode-book.com/}.
\bibitem[{Alfonso et~al.(2024)Alfonso, Conrardy, Sulejmani, Nirumand, Ul~Haq, Gomez-Vazquez, Sottet, and Cabot}]{alfonso2024building}
\bibinfo{author}{I.~Alfonso}, \bibinfo{author}{A.~Conrardy}, \bibinfo{author}{A.~Sulejmani}, \bibinfo{author}{A.~Nirumand}, \bibinfo{author}{F.~Ul~Haq}, \bibinfo{author}{M.~Gomez-Vazquez}, \bibinfo{author}{J.-S. Sottet}, \bibinfo{author}{J.~Cabot},
\newblock \bibinfo{title}{Building besser: an open-source low-code platform},
\newblock in: \bibinfo{booktitle}{International Conference on Business Process Modeling, Development and Support}, \bibinfo{organization}{Springer}, \bibinfo{year}{2024}, pp. \bibinfo{pages}{203--212}.
\bibitem[{Cabot and Claris{\'{o}}(2023)}]{CabotC23}
\bibinfo{author}{J.~Cabot}, \bibinfo{author}{R.~Claris{\'{o}}},
\newblock \bibinfo{title}{Low code for smart software development},
\newblock \bibinfo{journal}{{IEEE} Softw.} \bibinfo{volume}{40} (\bibinfo{year}{2023}) \bibinfo{pages}{89--93}.

\end{thebibliography}


\end{document}